\documentclass[12pt,preprint]{aastex}

\newcommand{\be}{\begin{equation}}
\newcommand{\ee}{\end{equation}} 
\newcommand{\simless}{\lower.5ex\hbox{$\; \buildrel < \over \sim\;$}}
\newcommand{\simgreat}{\lower.5ex\hbox{$\; \buildrel > \over \sim\;$}} 
 
\newcommand{\rclust}{{ R_{\rm c} }} 
 
\newcommand{\xrayexp}{{\langle L_{X} \rangle_\ast }} 
\newcommand{\xraycluster}{{\langle L_{X} \rangle_C }}

\newcommand{\barlumx}{ \bar{L}_X } 

\begin{document}

\title{Background X-ray Radiation Fields Produced by \\
Young Embedded Star Clusters} 

\author{Fred C. Adams$^{1,2}$, Marco Fatuzzo$^3$, and Lisa Holden$^4$, }
 
\affil{$^1$Physics Department, University of Michigan, Ann Arbor, MI 48109}   

\affil{$^2$Astronomy Department, University of Michigan, Ann Arbor, MI 48109}

\affil{$^3$Physics Department, Xavier University, Cincinnati, OH 45255}

\affil{$^4$Department of Mathematics, Northern Kentucky University, Highland Heights, KY 41099} 

\email{fca@umich.edu, fatuzzo@xavier.edu, holdenl@nku.edu} 

\begin{abstract} 

Most star formation in our galaxy occurs within embedded clusters, and
these background environments can affect the star and planet formation
processes occurring within them.  In turn, young stellar members can
shape the background environment and thereby provide a feedback
mechanism. This work explores one aspect of stellar feedback by
quantifying the background X-ray radiation fields produced by young
stellar objects.  Specifically, the distributions of X-ray
luminosities and X-ray fluxes produced by cluster environments are
constructed as a function of cluster membership size $N$. Composite
flux distributions, for given distributions of cluster sizes $N$, are
also constructed. The resulting distributions are wide and the X-ray
radiation fields are moderately intense, with the expected flux levels
exceeding the cosmic and galactic X-ray backgrounds by factors of
$\sim10-1000$ (for energies 0.2 -- 15 keV).  For circumstellar disks
that are geometrically thin and optically thick, the X-ray flux from
the background cluster dominates that provided by a typical central
star in the outer disk where $r \ga 9 - 14$ AU.  In addition, the
expectation value of the ionization rate provided by the cluster X-ray
background is $\zeta_X\sim8\times10^{-17}$ s$^{-1}$, about 4 -- 8
times larger than the canonical value of the ionization rate from
cosmic rays.  These elevated flux levels in clusters indicate that
X-rays can affect ionization, chemistry, and heating in circumstellar
disks and in the material between young stellar objects.

\end{abstract}


\section{Introduction} 

Most star formation in our galaxy occurs within embedded clusters,
which are themselves located inside giant molecular clouds.  The
radiation fields produced by these stellar nurseries can influence the
formation of additional cluster members, and especially their
accompanying planetary systems (Adams 2010). More specifically, the
processes of star and planet formation can be affected via [1] heating
of starless cores, leading to evaporation and the loss of star forming
potential (e.g., Gorti \& Hollenbach 2002), [2] ionization within
starless cores, leading to greater coupling between the magnetic
fields and gas (e.g., Shu 1992), thereby acting to suppress continued
star formation, [3] evaporation of circumstellar disks, leading to
loss of planet forming potential (e.g., Shu et al. 1993, Hollenbach et
al. 1994, St{\"o}rzer \& Hollenbach 1999, Adams et al. 2004), and [4]
ionization of circumstellar disks, which helps maintain the
magneto-rotational instability (MRI), which in turn helps drive disk
accretion (e.g., Balbus \& Hawley 1991).  With regard to circumstellar
disks, we note that the background radiation from the cluster
environment often dominates that produced by the central star,
especially at ultraviolet (UV) wavelengths (e.g., Hollenbach et al.
2000; Armitage 2000; Adams \& Myers 2001, Fatuzzo \& Adams 2008);
nonetheless, the photoevaporation of disks is often driven by stellar
radiation (e.g., Shu et al. 1993; Alexander et al. 2004, 2005, 2006).

Multi-wavelength observations of embedded clusters over the past two
decades have provided a wealth of information on these environments
(Lada \& Lada 2003; Porras et al. 2003; Allen et al. 2007), making it
possible to assess what impact a given physical process can have on
star and planet formation. Indeed, a comprehensive analysis of the
effects of FUV and EUV background fields on cluster environments has
already been performed (Armitage 2000; Adams et al. 2006; Fatuzzo \&
Adams 2008; Hollenbach \& Gorti 2009; Holden et al. 2010). The main
goal of this paper is to perform a similar analysis for the X-ray
band.

This paper is organized as follows. In Section 2, we discuss the
cluster environment, present our characterization of the stellar IMF,
and outline the basic approach for assigning X-ray luminosities to
stars in our cluster models.  We discuss the statistical aspects of
our cluster model in Section 3 and present distributions of the X-ray
luminosities for clusters with varying stellar membership size $N$.
In Section 4, we construct the corresponding distributions of X-ray
fluxes, both for varying $N$ and for different forms for the stellar
density profiles. The implications of these findings are discussed in
Section 5; the paper concludes in Section 6 with a summary of results
and a discussion of potential applications.

\section{The Cluster Environment}

Studies of clusters out to 2 kpc (Lada \& Lada 2003) and out to 1 kpc
(Porras et al. 2003) indicate that in the solar neighborhood, the
number of stars born in clusters with $N$ members is (almost) evenly
distributed logarithmically over the range $N \approx$ 30 to 3000,
with half of all stars belonging to clusters with $N \la 300$. These
stars are contained within a cluster radius $R_{c}$ ranging between
$0.1 - 2$ pc, with the radial sizes of observed clusters following an
empirically  determined law of the form 
\be 
\rclust (N) = 1\,{\rm pc}\,\left({N \over 300}\right)^\gamma \,,
\label{rofn}
\ee 
with $\gamma = 1/2$ (see Figure 2 of Adams et al. 2006, which
uses the data from Carpenter 2000 and Lada \& Lada 2003).  For
simplicity, we adopt this relation to specify the cluster radius
throughout most of this paper, although we use $\gamma = 1/3$
when considering a population of clusters that extend up to $N = 10^5$
in order to be consistent with the radial sizes observed for larger
clusters (for further discussion and supporting data, see Chandar et
al. 1999; Pfalzner 2009; Proszkow \& Adams 2009).  The gas within the
cluster environment not used to build stars accounts for roughly 70 --
90\% of the total (gas and stellar) mass of the cluster, and extends
well beyond $R_c$ until it eventually merges smoothly into the
background of the molecular cloud.
 
We assume that the stars within a cluster are drawn from a parent
population that is described by the universal initial mass function
(IMF) observed in our Galaxy.  The resulting probability distribution
is therefore independent of membership size $N$.  We adopt the broken
power-law form of the IMF presented in Kroupa (2001; not corrected for
binaries), expressed in terms of dimensionless mass $m = M/M_\odot$
and truncated at $m = 100$, so that
\be
{dN_\star\over dm} = C_i\, m^{-\alpha_i}\,,
\label{imf} 
\ee
where
\begin{eqnarray}
\alpha_0 = 0.3 & \qquad 0.01\le m \le 0.08\,, \nonumber \\
\alpha_1 = 1.3 & \qquad 0.08\le m \le 0.5\,, \\
\alpha_2 = 2.3 & \qquad 0.5\le m \le 100\,. \nonumber 
\label{imf2} 
\end{eqnarray}
The constants $C_i$ are
easily determined through the continuity of the IMF and the
normalization of equation (\ref{imf}) to the number of stars within a
cluster.  For completeness, we note that the expectation value of the
stellar mass for the adopted IMF is given by 
\be
\langle m \rangle \equiv {1\over N_\star}  
\int_{0.01}^{100}  m \, \left({dN_\star\over dm}\right) \, dm = 0.38\,,
\label{meanmass} 
\ee
where $N_\star$ = $\int(dN_\star/dm)dm$. 

\begin{figure}
\figurenum{1}
{\centerline{\epsscale{0.90} \plotone{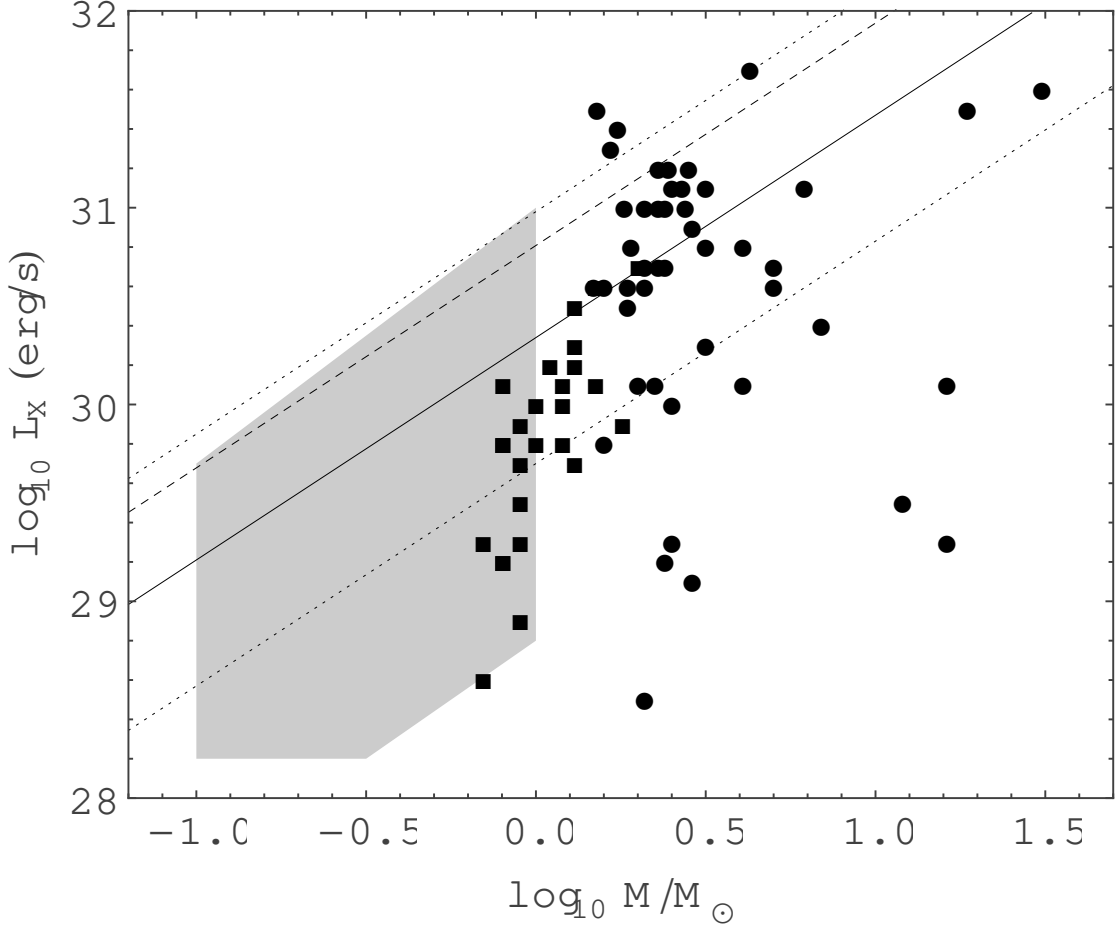} }}
\figcaption{Summary of observational data for the mass vs X-ray  
luminosity relation for stars in young cluster environments.  Squares
-- low mass Chamaeleon I stars presented in Table 6 of Feigelson et
al. (1993); circles -- intermediate/high mass ONC stars presented in
Table 5 of Feigelson et al. (2002); grey region -- region of parameter
space spanned by soft X-ray emission of low mass ONC stars, as
presented in Figure 14 of Feigelson et al. (2002).  The solid line
shows the results of the linear regression fit obtained by Preibish et
al. (2005) using the PS model, and the dotted lines mark the
corresponding $\pm w$ deviations, where $w = \Delta (\log L_X)$ is the
residual standard deviation.  The dashed line represents the mean
value $\barlumx (m)$ as given by equation (\ref{fxraylum}) in the
text.}
\label{fig:xraylum} 
\end{figure}

We now consider the X-ray luminosity of the stars within an embedded
cluster. X-rays from low-mass stars ($m \la 2$) are produced from
flaring events in stellar coronae, whereas X-ray emission from
high-mass stars ($m \ga 18$) is driven by stellar winds (intermediate
mass stars most likely do not produce X-rays, with detections from
such stars then attributed to low-mass companions).  It is not
surprising, therefore, that observations of X-ray luminosities of
cluster members show a large variance in X-ray values (e.g, Feigelson
et al. 1993, Feigelson et al. 2002, Preibisch et al. 2005); the
variance also tends to decrease with the cluster age (see Alexander \&
Preibisch 2012). Nonetheless, these observations indicate that the
X-ray luminosity depends on stellar mass. Modeling the exact nature of
this dependence is hampered by the large variance in the data and by
the fact that the derived correlation between $L_X$ and $m$ is model
dependent (Preibisch 2005).  In our work, we adopt a purely empirical
approach and use a monotonic relation between stellar mass and X-ray
luminosity.  Specifically, we use the results obtained by Preibish et
al. (2005), and focus primarily on the PS model for which a linear
regression fit yields the relation 
\be 
\log\left(L_X[\rm{erg\,s}^{-1}] \right) = 30.34 + 1.13 \, \log m \;,
\label{lxmass_ps} 
\ee
and a standard deviation of $\Delta (\log L_X) = 0.64$.  As shown in
Figure \ref{fig:xraylum}, this best fit line (solid line) and the
corresponding $\pm \Delta (\log L_X)$ deviations (dotted lines) are in
reasonable agreement with the low-mass data (squares) obtained from
Chamaeleon I (Feigelson et al. 1993) and the low-mass data (gray area)
and intermediated/high-mass data (circles) obtained from the Orion
Nebula (Feigelson et al. 2002).  For completeness, we also calculate
cluster luminosity distributions using results from the SDF models,
for which a linear regression fit yields the relation
\be
\log\left(L_X[\rm{erg\, s}^{-1}] \right) = 30.37 + 1.44 \, \log m \;,
\label{lxmass_sdf} 
\ee
and a standard deviation of $\Delta (\log L_X) = 0.65$.  

As elaborated on in Section 3, we obtain X-ray luminosity
distributions for clusters with $N$ members by randomly sampling the
IMF for each stellar member, and then assigning a luminosity to each
star based on its mass. For a fixed stellar mass, we include a
distribution of X-ray luminosities; guided by the aforementioned
observations, this latter step is performed by randomly selecting a
value of $\log\left[L_X(m)\right]$ from a normal distribution with the
mean given by equation (\ref{lxmass_ps}) or equation
(\ref{lxmass_sdf}) and with a standard deviation $w$
$\equiv\Delta(\log L_X)$. We note that for the PS model, the mean
luminosity of the resulting distribution for stars of mass $m$ is 
\be
\barlumx (m) \,[{\rm erg \,s}^{-1}]= \int_{-\infty}^{\infty} 
P[x; 30.34 +1.13 \log m, 0.64] 10^x dx =  6.4 \times 10^{30} \,m^{1.13}\;,
\label{fxraylum}
\ee
where $P[x;\mu,w]$ is the probability density function of mean $\mu$
and standard deviation $w$.  This value is denoted by the dashed line
in Figure \ref{fig:xraylum}.  Since the values of 
$\log\left[L_X(m)\right]$ are selected from a normal distribution, the
corresponding distribution of $L_X$ values are not normal, but rather,
are positively skewed.  As a result, $\log \barlumx (m) > 30.34 + 1.13
\log m$ (as seen in Figure 1).  The expectation value of the X-ray
luminosity --- per star --- over the full distribution of stellar
masses is then given by 
\be
\xrayexp \equiv {1\over N}  \int_{0.01}^{100} 
\barlumx (m)\,{dN_\star\over dm} \,dm = 
2.6 \times 10^{30}\;{\rm erg \,s}^{-1}.
\label{lxmean} 
\ee 

\begin{figure}
\figurenum{2}
{\centerline{\epsscale{0.90} \plotone{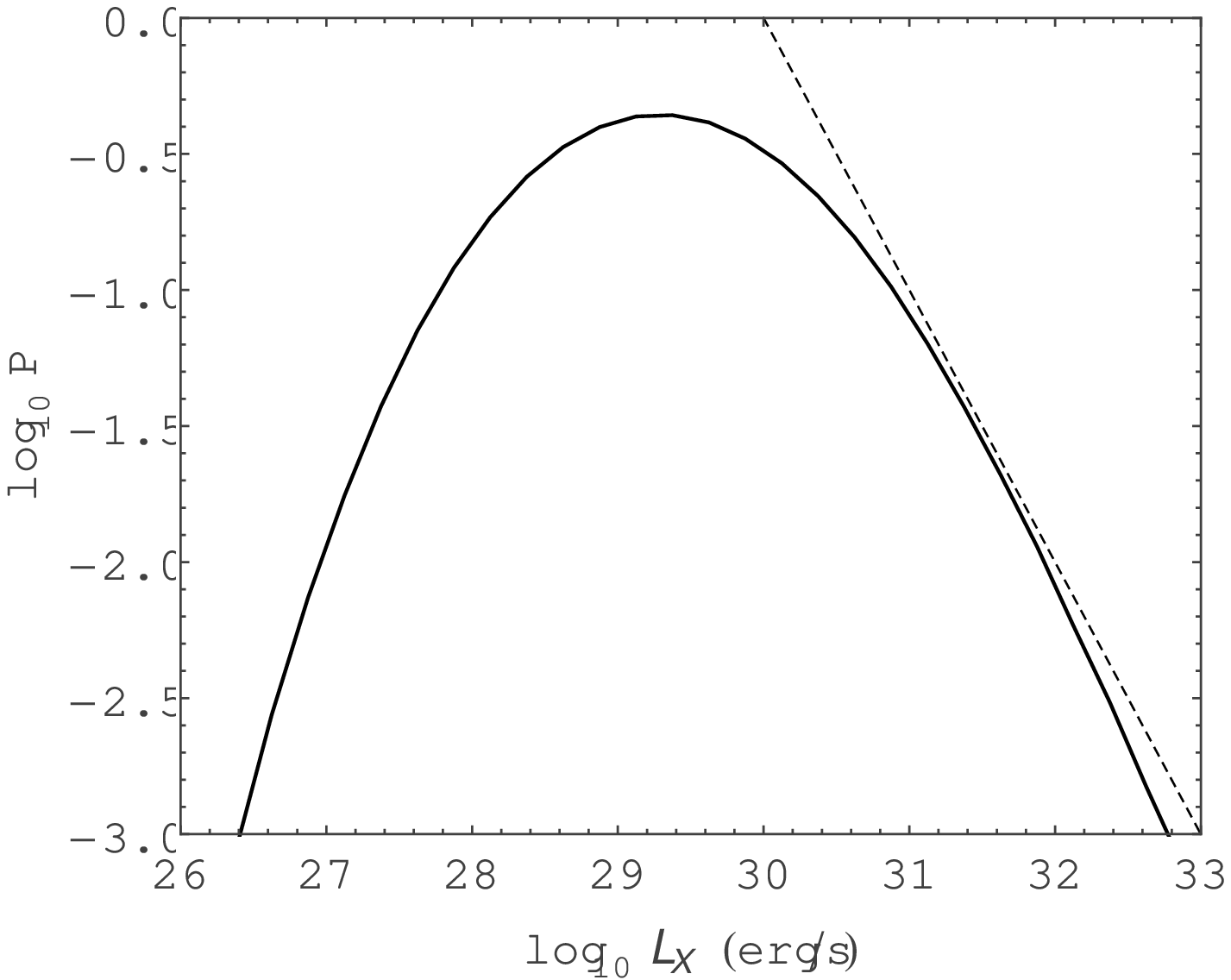} }}
\figcaption{Probability distribution of the X-ray luminosities. 
This distribution was obtained by first randomly selecting a stellar
mass from the IMF defined by equations (\ref{imf}) and (\ref{imf2}),
and then obtaining the corresponding luminosity by randomly selecting 
$L_X$ from a normal distribution of $\log L_X$ with mean given by 
equation (\ref{lxmass_ps}) and standard deviation $w = \Delta 
(\log L_X) = 0.64$ as discussed in the text.  The dotted line has 
slope --1 and is included for illustrative purpose. }
\label{fig:probability}
\end{figure}

We next consider the contribution to the total X-ray luminosity that
comes from stars with mass $m$ for a parent population drawn from our
adopted IMF.  We start by noting that the function 
\be 
\psi_X (m) \equiv \barlumx(m)\, {dN_\star \over dm} \, , 
\ee 
which characterizes the X-ray luminosity produced by the stellar
population (as a function of mass), peaks at $m = 0.08$ and drops off
as a power-law $\psi_X \propto m^{-1.17}$ for masses $m > 0.5$. As a
result, most of the luminosity of a cluster per mass decade, which
scales as $m\psi_X$, is expected to come from stars with mass $m \ga
0.5$.  To illustrate this point, we randomly select $10^7$ stars from
our IMF and then select the corresponding luminosity from equation
(\ref{lxmass_ps}), including the variance about this mean relationship
as prescribed above.  The resulting X-ray luminosity distribution is
shown in Figure \ref{fig:probability}.  The distribution displays a
broad peak near $L_X \sim 10^{29}$ erg s$^{-1}$ $\approx\barlumx
(0.08)$, and approaches a slope of $-1$ (as indicated by the dotted
line) as $L_X \rightarrow 10^{31}$ erg s$^{-1}$ $\sim \barlumx (0.5)$.
The distribution of X-ray luminosity per mass decade at higher values
is therefore nearly flat, indicating that above $m\approx 0.5$, stars
ranging in mass between $m = 10 \rightarrow 100$, although fewer in
number, can contribute nearly the same to the total cluster X-ray
luminosity as stars ranging in mass between $m = 1\rightarrow 10$.
This result is consistent with the weak dependence of the function
$m\psi_X \propto m^{-0.17}$ on mass above $m=0.5$.  It is worth noting
that our calculated distribution, as shown in Figure 2, is in good
agreement with the observed distribution of X-ray luminosities of the
Orion Nebula population (see Figure 3a in Feigelson et al. 2005).

\section{X-Ray Luminosity Distributions} 

With the IMF and the prescription for obtaining the luminosity of the
sources specified, we now determine the characteristics of the X-ray
luminosity distribution for a given cluster of size $N$, including
both the expectation value and the variance. The X-ray luminosity for
a single cluster, denoted as $L_X(N)$, is given by the sum
\be
L_X (N) = \sum_{j=1}^N L_{Xj} \, , 
\label{uvsum} 
\ee
where $L_{Xj}$ is the X-ray luminosity from the $jth$ member, and $N$
is the membership size of the cluster.  In this formulation, we assume
that the X-ray luminosity for a given star is determined by a
distribution whose properties depend only on the stellar mass, and
that the stellar mass is drawn independently from the specified
stellar IMF. The sum in equation (\ref{uvsum}) is thus the sum of
random variables, where the variables are drawn from known and
independent distributions (which are determined by the IMF and the
$L_{X}-m$ relation).  In the limit of large $N$, the expectation value
$\xraycluster$ of the X-ray power for the cluster (which is a function 
of $N$) is given by the expression 
\be 
\xraycluster = N \xrayexp \, ,  
\label{lnexp} 
\ee 
where $\xrayexp$ is the expectation value of the X-ray power per star,
as given by equation (\ref{lxmean}).  As usual, the central limit
theorem implies that the distribution of values $L_{X}(N)$ obtained by
sampling over many clusters must approach a Gaussian form in the limit
$N \to \infty$ (e.g., Richtmyer 1978), although convergence is often
slow. One of the issues of interest here is the value of stellar
membership $N$ required for these statistical considerations to be
valid; similarly, we would like to know the fraction of the cluster
population that has such sufficiently large $N$. In its limit of
applicability, this Gaussian form for the composite distribution is
independent of the form of the initial distributions, i.e., the shape
of the distribution is independent of the stellar IMF and the
mass-luminosity relation. The width of the distribution also converges
to a known value given by the expression
\be 
\langle w \rangle_X^2 = {1 \over N} \sum_{j=1}^N w_j^2 \quad
\Rightarrow \quad\langle w \rangle_X= \sqrt{N} w_0 \, ,
\label{nvariance} 
\ee 
where $w_0$ is the width of the individual distribution, i.e., 
\be 
w_0^2 \equiv \langle L_X^2 \rangle_\ast - \xrayexp^2 \, .  
\ee 
We can write $w_0=Q\xrayexp$, where $Q$ is the dimensionless width of
the X-ray luminosity distribution. Using the specifications given
above, we find $Q\approx17$.  For a given membership size $N$, the
mean value of the distribution of cluster luminosities converges to
the value $N\langle{L_X}\rangle_\ast$, whereas the width of the
distribution converges to $\sqrt{N}Q\langle{L_X}\rangle_\ast$. 
As a result, the mean value will be larger than the width of the
distribution only if the number of stellar members exceeds
$N_Q=Q^2\approx300$. Small clusters (with $N<N_Q\approx300$) thus have
such wide distributions of X-ray luminosity that the mean value does
not provide a good estimate. For larger clusters (with $N>N_Q$), the
X-ray luminosity is adequately characterized by the expectation value,
but the distributions retain significant (relative) widths. These 
trends are illustrated in the figures that follow below. 

\begin{figure}
\figurenum{3}
{\centerline{\epsscale{0.90} \plotone{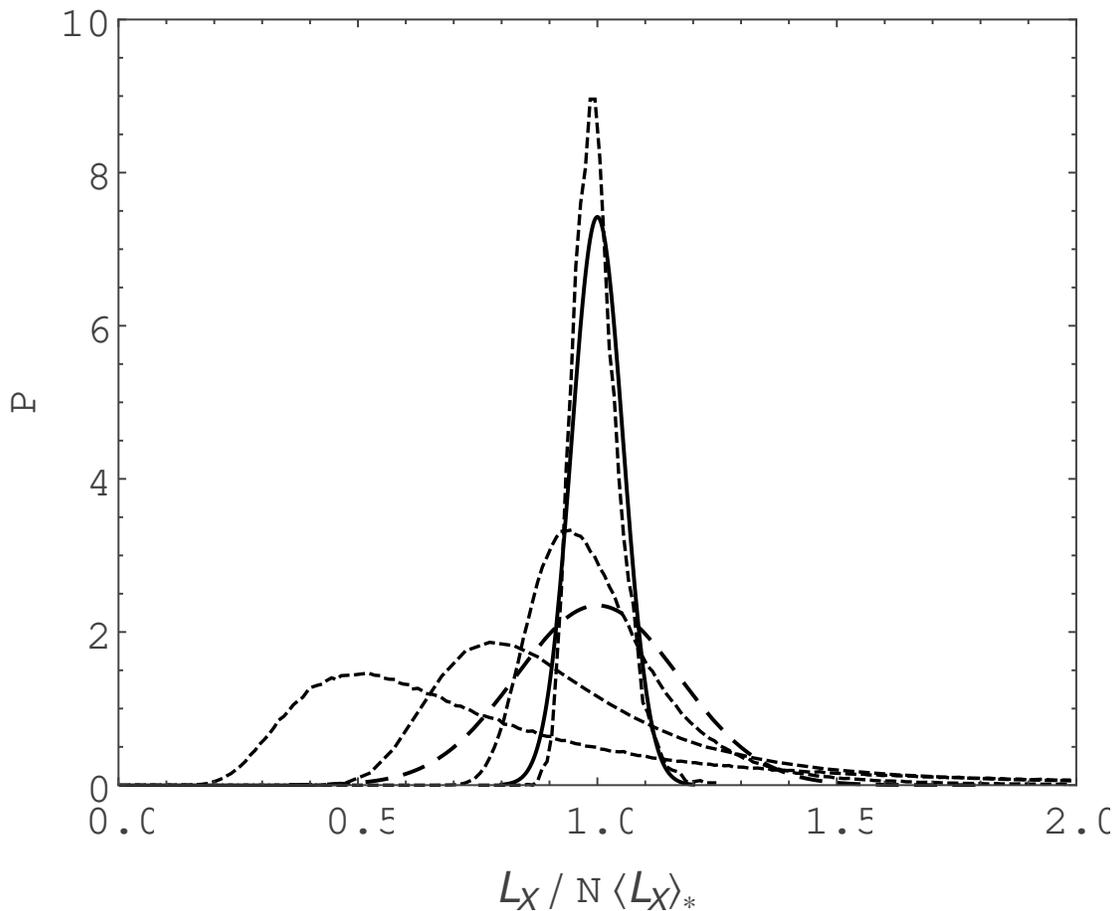} }}
\figcaption{Distribution of X-ray luminosities. The short-dashed curves
show the probability distributions for clusters with (from left to right)  
$N = 10^2$ stars, $N$ = $10^3$ stars, $N = 10^4$ stars, and $N = 10^5$
stars. Note that the luminosities (along the horizontal axis) are
normalized by the average value one would obtain with complete sampling
of the stellar IMF.  For comparison, the long-dashed curve shows a
Gaussian distribution with mean $\mu = 1$ and standard deviation 
$w=Q/\sqrt{10^4}$, and the solid curve shows a Gaussian distribution
with mean $\mu = 1$ and standard deviation $w=Q/\sqrt{10^5}$. }
\label{fig:lumdist} 
\end{figure}

Figure \ref{fig:lumdist} illustrates the X-ray luminosity
distributions obtained from our cluster model, scaled by the cluster
expectation value $N \xrayexp$ from equation (\ref{lnexp}), for
clusters with membership size $N=10^2$, $10^3$, $10^4$, and $10^5$
(short-dashed curves, from left to right).  The long-dashed curve
represents a Gaussian profile with mean $\mu = 1$ and standard
deviation $w = Q/\sqrt{10^4}$, whereas the solid curve represents a
Gaussian profile with mean $\mu = 1$ and standard deviation $w =
Q/\sqrt{10^5}$.  It is clear from this result that as $N$ increases,
the sampling becomes more complete, and the distributions approach a
Gaussian form. However, in practice, even with large values of cluster
membership size $N$, the asymptotic form is not reached.  We note that
the clusters in our solar neighborhood, as observed by Lada \& Lada
(2003) and Porras et al. (2003), are in the regime of incomplete
sampling, and the distributions of their X-ray luminosities are
expected to have median values that are smaller than $N \xrayexp$
while retaining significant tails.

\begin{figure}
\figurenum{4}
{\centerline{\epsscale{0.90} \plotone{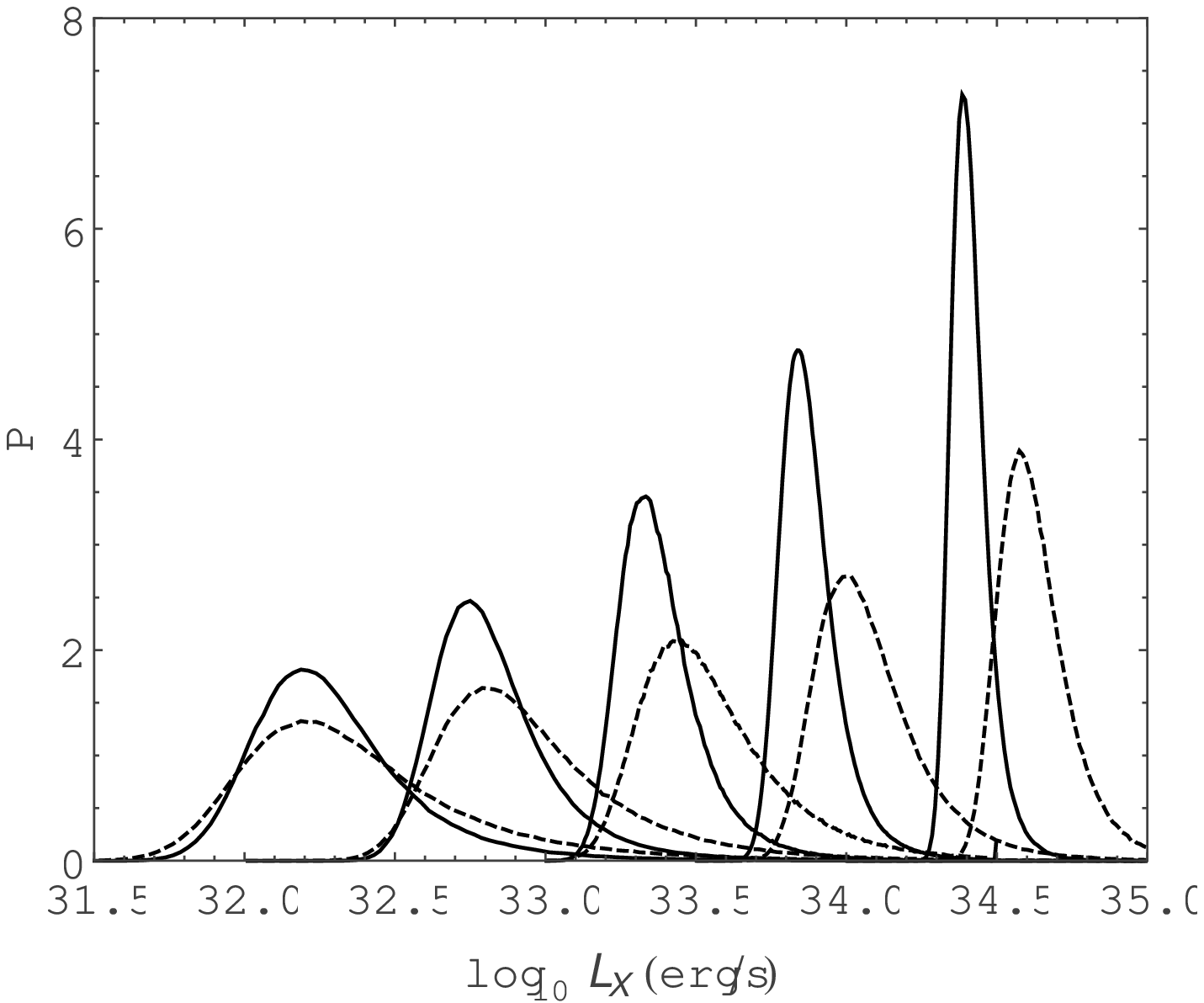} }}
\figcaption{Probability distribution for X-ray luminosities. The five 
pairs of curves show the probability distributions for clusters of
size (from left to right) $N$ = 100, 300, 1000, 3000 and 10000. Solid
curves were obtained using the PS model; dashed curves were obtained
using the SDF model. }
\label{fig:lumdistalt} 
\end{figure}

Figure \ref{fig:lumdistalt} shows the X-ray luminosity distributions
(with physical units) for clusters with stellar membership size $N$ =
100, 300, 1000, 3000 and 10000, constructed using both the PS model
(solid curves) and the SDF model (dashed curves) for the X-ray
luminosity vs mass relation.  As expected, the mean/median values of
the distributions increase with increasing $N$; in addition, the width
of the distribution decreases with increasing $N$. The SDF models
produce distributions with larger means and standard deviations (by
about a factor of 2 for larger clusters) than their PS model
counterparts.  These characteristics result from the stronger mass
dependence for the luminosity in the SDF models, resulting in a
broader range of luminosity values sampled in that case.

\section{X-Ray Flux Distributions} 

In this section we construct distributions of the X-ray flux impinging
upon the cluster members.  Unlike the FUV and EUV bands, for which a
significant fraction of the luminosity is produced by the most massive
member of the cluster (e.g., Fatuzzo \& Adams 2008), the total X-ray
luminosity can arise from comparable contributions of numerous stars.
As a result, the overall X-ray emission can be spread out over the
entire cluster region, although this assumption is not universally
true.  For example, about 50\% of the total X-ray luminosity from all
($N\ga1500$) of the X-ray emitting stars in the Orion Nebula Cluster
comes from the most massive star. Nevertheless, we can obtain a
benchmark estimate for the X-ray flux within the cluster environment
by dividing the expectation value of the luminosity $N \xrayexp$ by
the fiducial surface area $A$ = $\pi R_c(N)^2$, where $R_c(N)$ is
given by equation (\ref{rofn}), which yields a value of 
\be 
F_{X0} = {N \xrayexp \over \pi R_c(N)^2} =
2.6 \times 10^{-5}\, {\rm erg} \, {\rm cm}^{-2} \, {\rm s}^{-1} \;.
\label{benchmark} 
\ee
  
Of course, the X-ray fluxes within a cluster, and within a collection
of clusters, will take on a distribution of values, where $F_{X0}$
represents an order-of-magnitude estimate for the mean.  In order to
construct these distributions of the X-ray flux, we first randomly
select the position of a test star within a radius $R_c(N)$ (assuming
a spherical distribution), weighted by a chosen density profile.  For
this study, we consider density profiles with the simple forms
$\rho\propto1/r$ and $\rho\propto1/r^2$; these profiles roughly
bracket the possible profiles expected from observations.  The
probability that the test star is located a distance between $r$ and
$r+dr$ from the cluster center is then given by  
\be
P(r)\,dr = {\rho\, r^2 dr \over \int_0^{R_c} \rho \,r^2 dr}\;.
\ee
We then randomly selected the positions of the remaining $N-1$ stars
within the cluster, following the same prescription used to set the
location of the test star.  For each of these field stars, a
luminosity is obtained as detailed in Section 2. The flux impinging
on the test star is calculated, and the entire process is then
repeated $10^7$ times in order to build up flux distributions.  Note
that this procedure builds a slightly different distribution than the
one that describes the flux values found within the cluster
environment, for which case the location of the test star (which was
weighted in accordance to the specified density profile) would have
been replaced by a randomly selected location within the cluster that
was not weighted by the density profile.

\begin{figure}
\figurenum{5}
{\centerline{\epsscale{0.90} \plotone{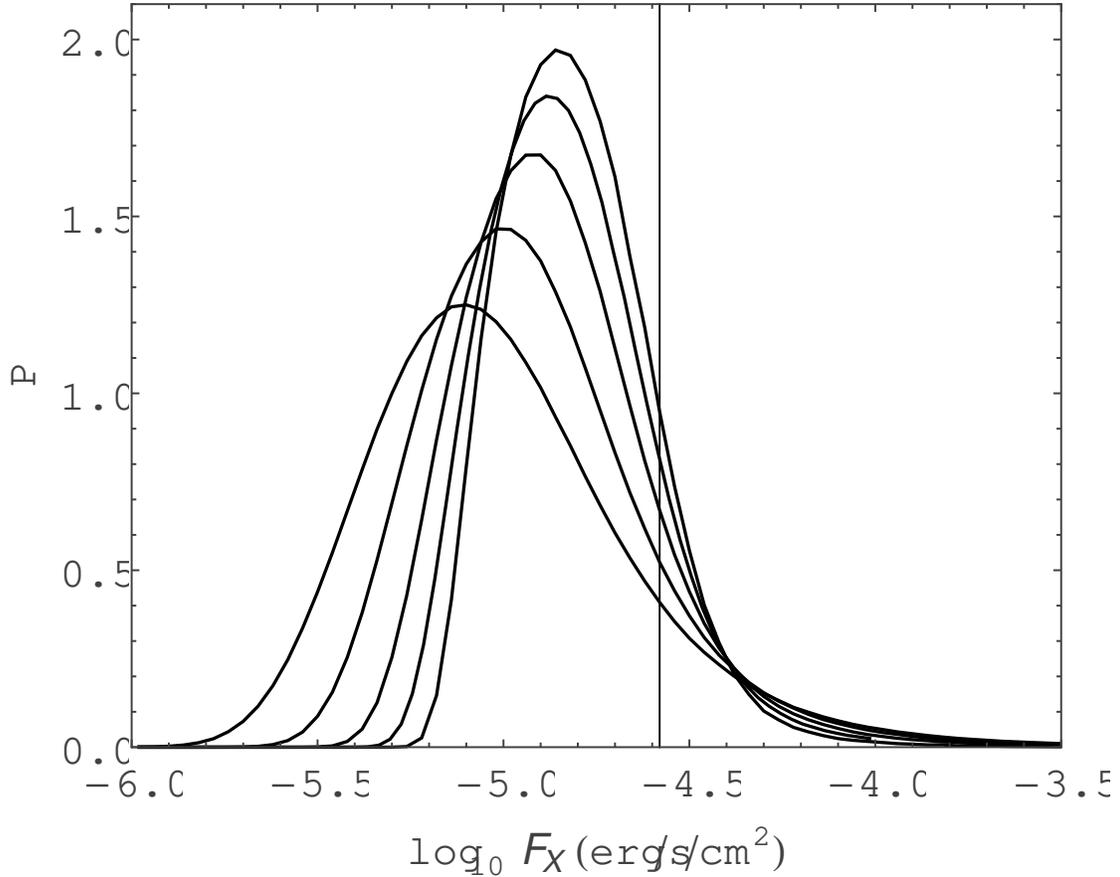} }}
\figcaption{Probability distribution for the X-ray flux impinging upon   
stellar members for clusters where the stars follow a radial density
distribution $\rho\propto1/r$.  The five curves show the probability
distributions (left to right) for clusters with varying membership
sizes $N$ = 100, 300, 1000, 3000 and 10000.  The vertical line denotes
our benchmark value of X-ray flux defined by equation (\ref{benchmark}). } 
\label{fig:fluxdist} 
\end{figure}

\begin{figure}
\figurenum{6}
{\centerline{\epsscale{0.90} \plotone{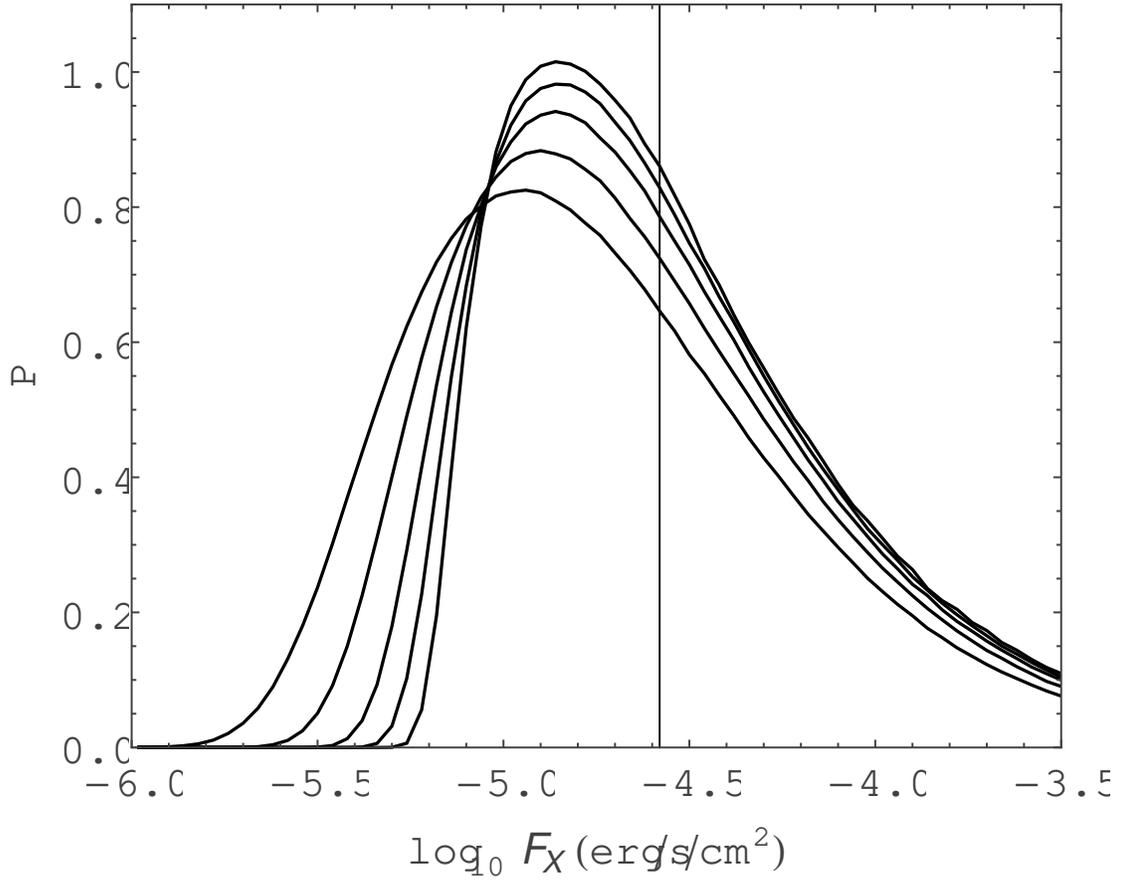} }}
\figcaption{Same as Figure \ref{fig:fluxdist}, but for stars 
following a radial density distribution with $\rho\propto 1/r^2$.}
\label{fig:fluxdistalt} 
\end{figure}

The results are shown in Figure \ref{fig:fluxdist} (for $\rho\sim1/r$)
and in Figure \ref{fig:fluxdistalt} (for $\rho \sim 1/r^2$). The range
of fluxes realized within the clusters is comparable to (but generally
less than) the benchmark estimate of equation (\ref{benchmark}).  Note
that the flux distributions peak at nearly the same locations for both
density profiles (we use the same expression to set the cluster radius
for both cases, so that the mean stellar density for a given cluster
size $N$ is the same for both density profiles).  However, stars with
a $\rho\propto 1/r^2$ profile are more centrally localized, leading to
a distribution that extends out to higher values of flux. As the
stellar membership size $N$ increases, sampling of the stellar IMF
becomes more complete, and the flux distributions shift to higher
values (they move to the right in the figures) and become somewhat
narrower. In rough terms, however, the expected X-ray flux levels are
given by $F_X \sim 10^{-5}$ erg cm$^{-2}$ s$^{-1}$, with a factor of
$\sim3 - 4$ variation for $\rho\propto 1/r$ profiles and $\sim5 - 10$
variations (above the peak) for $\rho\propto 1/r^2$ profiles.

\begin{figure}
\figurenum{7}
{\centerline{\epsscale{0.90} \plotone{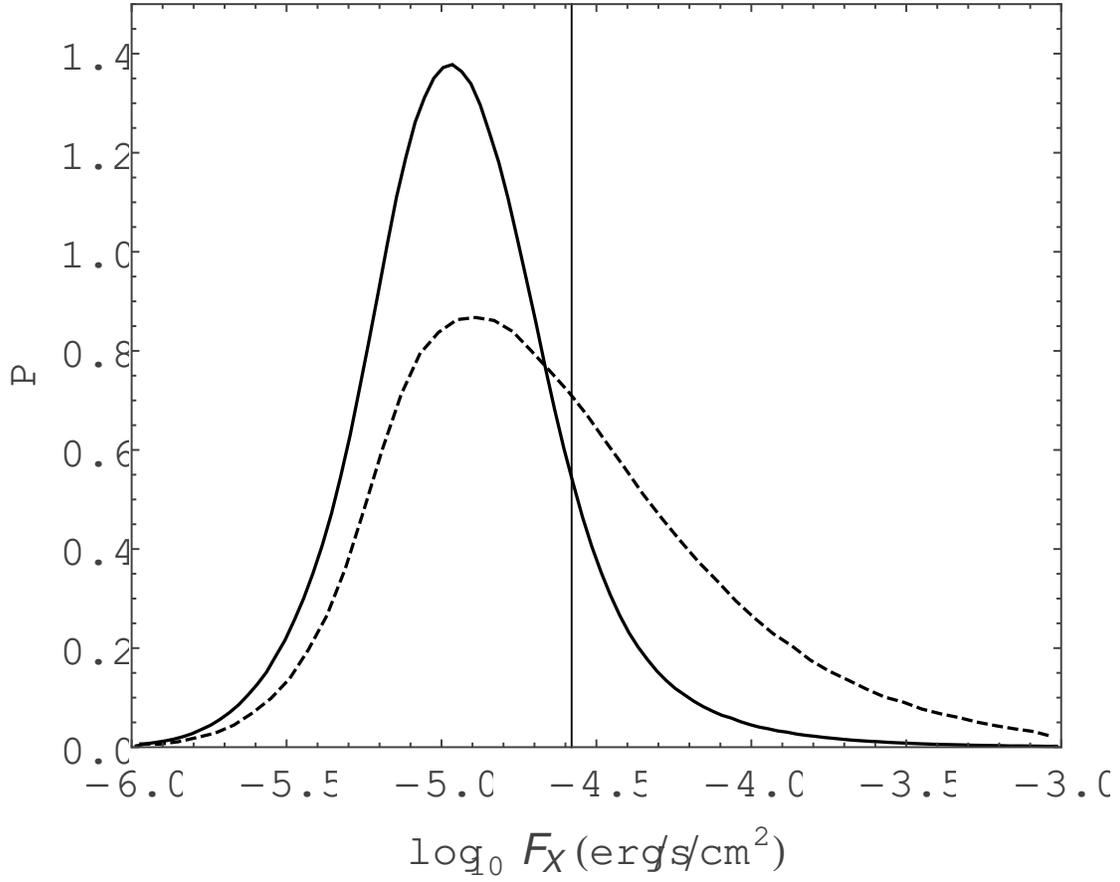} }}
\figcaption{Composite distribution for the X-ray flux impinging upon 
stellar members in a population of clusters similar to that observed
in the solar neighborhood with $N_{max} = 3000$.  The solid curve was
obtained by assuming a density profile of the form $\rho\propto 1/r$.
The dashed curve was obtained by assuming a density profile of the
form $\rho\propto 1/r^2$.  The vertical line denotes our benchmark
value of flux as defined by equation (\ref{benchmark}).}
\label{fig:flux_compnbd} 
\end{figure}

\begin{figure}
\figurenum{8}
{\centerline{\epsscale{0.90} \plotone{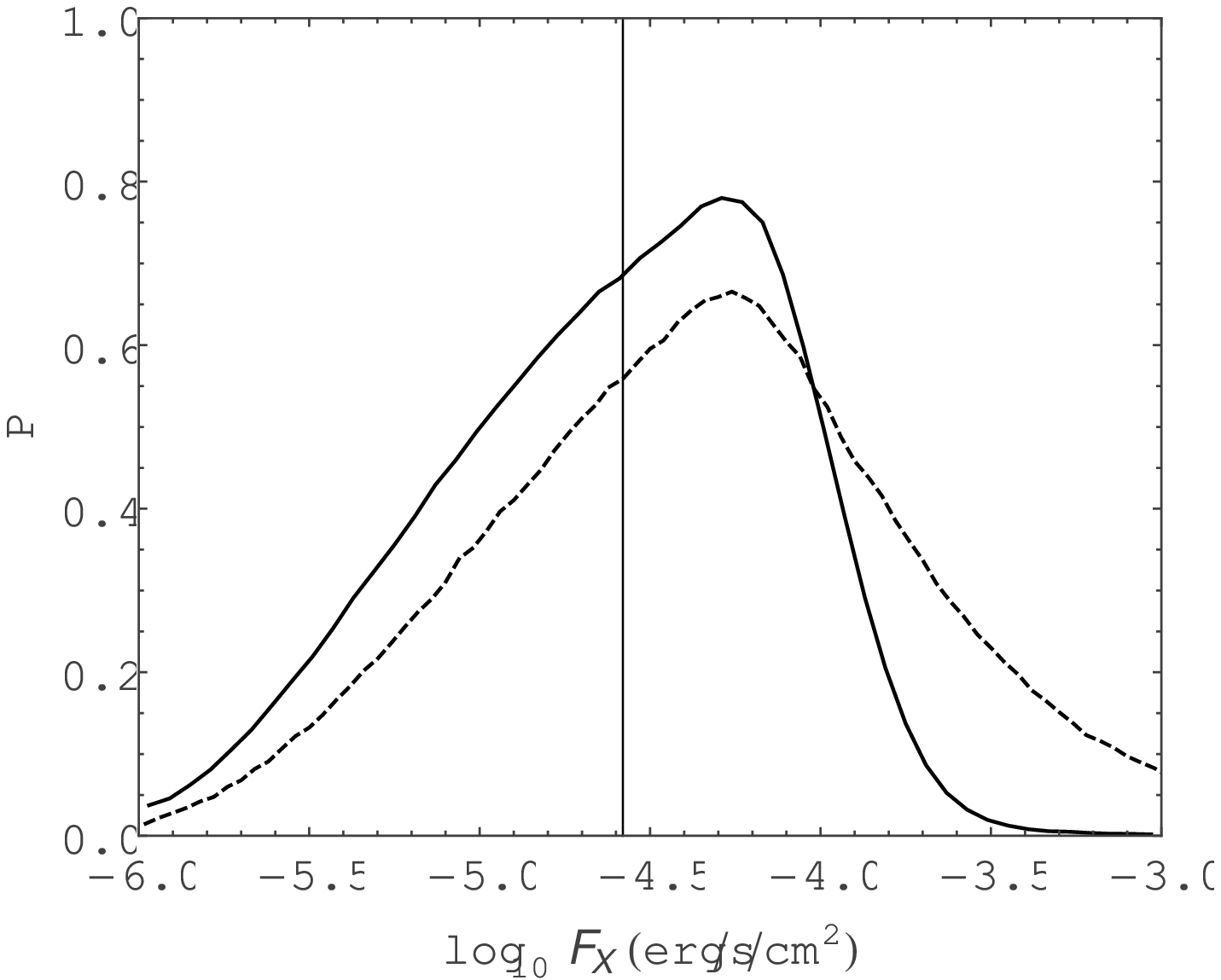} }}
\figcaption{Same as Figure 7, but for a population of clusters with 
membership size extending up to $N_{max} = 10^5$, and with radii given
by equation (\ref{rofn}) where $\gamma = 1/3$. }
\label{fig:flux_compbig} 
\end{figure}

We complete our analysis by constructing composite flux distributions
for all stars within: i) a cluster population similar to that observed
in our solar neighborhood, for which the largest cluster has $N_{max}$
= 3000 stars; and ii) a cluster population with cluster membership
size extending up to $N_{max} = 10^5$.  For case (i) we assume a
cluster radius as specified by equation (\ref{rofn}) with $\gamma$ =
1/2, whereas for case (ii) we assume $\gamma$ = 1/3.  Following the
observational result that clusters in the solar neighborhood have
sizes $N$ that are (almost) evenly distributed logarithmically, we
assume that both populations are described by the cumulative
distribution 
\be 
f(N) = {\log(N) - \log(30) \over \log(N_{max}) - \log(30)}\,, 
\ee 
where $f(N)$ denotes the fraction of stars born in clusters of stellar
membership less than or equal to $N$ (see, e.g., Figure 1 from Fatuzzo
\& Adams 2008).  Note that for $N_{max} = 3000$, $f(300) = 0.5$, so
that half of the stars in the entire cluster population belong to
clusters with $N \le 300$; for comparison, this ``median point'' is
much larger for the sample with $N_{max} = 10^5$, where $f(1732)$ =
0.5.  Composite distributions are then obtained by first sampling
over the assumed cluster population to set the size of the cluster
that our test star belongs to, and then performing the process
outlined above to calculate the flux impinging on that star as a
result of the other stellar members.  This process is then repeated 
$10^7$ times in order to build up a statistically valid distribution. 
The results are presented in Figures \ref{fig:flux_compnbd} and 
\ref{fig:flux_compbig}. 

\section{Implications and Discussion} 

With the X-ray flux distributions specified, we can now assess some of
the corresponding effects on the cluster and on forming stars (and
planetary systems).  We first compare the X-ray flux received by
circumstellar disks from the background cluster with that received
from their central stars.  If we ignore geometry, the sphere of
influence of the central star (in X-rays) is defined by the radius
where the X-ray flux provided by the background environment $F_{XC}$
of the cluster is equal to that provided by the star $F_{X\star}$,
i.e., 
\be 
F_{X\star} = {L_X \over 4 \pi r^2} = 0.039 \,\,{\rm erg}\,\,{\rm cm}^{-2} 
\,\,{\rm s}^{-1} \left( {L_X \over 10^{29} \,\,{\rm erg}\,\,{\rm s}^{-1} } \right) 
\left( {r \over 30 \,\,{\rm AU} } \right)^{-2} = F_{XC} \, , 
\label{fxest} 
\ee 
where we have scaled the X-ray luminosity to a typical value.  For
moderate-sized clusters with $N \le N_{max}=3000$ such as those found
in the solar neighborhood, and for $\rho\propto 1/r$ density profiles,
the X-ray flux provided by the background cluster has a characteristic
value $F_{XC}\approx2\times 10^{-5}$ erg cm$^{-2}$ s$^{-1}$, but
varies by factors of $3-4$ on either side of this benchmark (see
Figures \ref{fig:fluxdist} and \ref{fig:flux_compnbd}).  For clusters
with $\rho\propto 1/r^2$ density profiles, the X-ray flux has nearly
the same peak value, but varies by factors of $5-10$ (see Figure
\ref{fig:fluxdistalt}). As a result, the sphere of influence of the
star (in X-rays) extends out to about $r\approx1300$ AU, much larger
than the typical disk size. If we extend the cluster sample to larger
membership sizes with $N_{max}=10^5$, the characteristic value of the
X-ray flux increases to $F_{XC}\approx6\times10^{-5}$ erg cm$^{-2}$
s$^{-1}$ (see Figure \ref{fig:flux_compbig}) and the sphere of
influence of the star has radius $r\approx760$ AU, still somewhat
larger than most circumstellar disks.

Taken at face value, this finding would indicate that the X-ray flux
for the planet-forming region of the disk is (usually) dominated by
the central star rather than the background environment of the
cluster. However, this comparison ignores geometrical factors arising
because the disk is not perpendicular to the outgoing stellar
radiation.  The expected vertical extent $\Delta z$ of the disk, in
the context of X-ray absorption, is estimated to be $(\Delta z)/r
\approx 0.3 - 0.4$ (see Figure 1 of Gorti \& Hollenbach 2009). As a
result, the disks are relatively thin geometrically, and the
intercepted X-ray flux will be significantly smaller than the value
presented in equation (\ref{fxest}).  For the limiting case of a
geometrically flat disk, the flux intercepted by the disk has the
following radial dependence
\be
F_{X} (r) = {L_X \over 4 \pi R_\ast^2} {1 \over \pi} 
\left[ \arcsin u - u \left( 1 - u^2 \right)^{1/2} \right] 
\to {L_X \over 4 \pi R_\ast^2} {2 \over 3\pi} 
\left( {R_\ast \over r} \right)^3 \, , 
\label{xraystar} 
\ee
where $u \equiv R_\ast/r$ and where we have taken the large radius
limit to obtain the final expression (Adams \& Shu 1986). Using this
expression, we find that the central star dominates the X-ray flux
only out to $r \approx 14$ AU (9 AU) for clusters distributions with
$N_{max}$ = 3000 ($10^5$). For these systems, the outer disk thus
receives more X-ray flux from the cluster than from the central star.
As an example, consider radial position $r=30$ AU and the extended
cluster distribution with $N_{max}=10^5$; the X-ray flux from the star
(from equation [\ref{xraystar}]) is estimated as $F_X \approx$
$2\times10^{-6}$ erg cm$^{-2}$ s$^{-1}$, 30 times smaller than the
background radiation flux.

Although the background cluster can provide more X-ray flux to
circumstellar disks than their central stars, the X-ray irradiation of
the inner portions of the disks by coronal emission from young stars
remains important. Several studies, complementary to the analysis of
this paper, have addressed the process of photoevaporation driven by
X-rays from the central star (e.g., Ercolano et al. 2008, 2009). The
predicted mass loss rates can be substantial (typically of order
$10^{-8}$ $M_\odot$ yr$^{-1}$) and most of the mass loss takes place
at disk radii in the range 10 -- 20 AU, near the transition where
X-ray flux is dominated by the star (inside this radius) and the
background cluster (outside).

On the scale of the cluster itself, ionization of the cluster gas is
an important consideration. The main contribution to ionization in
star forming regions is often taken to be cosmic rays, which have a
fiducial ionization rate $\zeta_{CR} \approx 10^{-17}$ s$^{-1}$ 
(Shu 1992); more recent studies suggest a slightly higher value  
$\zeta_{CR} \approx 2.6 \times 10^{-17}$ s$^{-1}$ (van der Tak \& van
Dishoeck 2000). For comparison, the ionization rate due to the X-ray
background radiation is given by 
\be
\zeta_X = {F_X \over E_X} \sigma (E_X) {E_X \over \Delta \epsilon} 
J(\tau,x_0)\,,
\label{zetax} 
\ee
where $E_X \sim 1$ keV is the characteristic energy of
the X-rays, $\sigma(E_X)$ is the total photoelectic cross section
(see the discussion of Glassgold et al. 2000), the function $J$ is an
attenuation factor, which is, in turn, a function of the optical depth
$\tau$ in X-rays, and the threshold value $x_0$ = $E_0/kT_X$ (given by
the lowest energy $E_0$ of the X-rays under consideration and by the
temperature $T_X$ of the emission region). The function $J$ is
expected to be of order unity (Krolik \& Kallman 1983) and we will
take $J=1$ for simplicity. The remaining factor in equation
(\ref{zetax}) is the factor $E_X/(\Delta \epsilon)$, where $\Delta
\epsilon \approx$ 35 eV is the energy required to make an ion
pair. This ratio thus represents the enhancement factor resulting from
individual X-rays making multiple ionizations (note that
$E_X/(\Delta\epsilon)\sim30$). A power-law fit to the cross section 
results in the form 
\be
\sigma(E_X) = \sigma_0 (E_X/1 \,\,{\rm keV})^{-n} \, , 
\label{csection} 
\ee
where the index $n$ = 2.485 and the fiducial value of the cross
section $\sigma_0$ = $2.27 \times 10^{-22}$ cm$^2$ (Glassgold et al.
1997). Putting all of the above results together, we find that to
leading order the X-ray ionization rate becomes 
\be 
\zeta_X \approx 
8 \times 10^{-17} \,\, {\rm s}^{-1} \,\, 
\left( {F_{XC} \over 2\times10^{-5} \,\,{\rm erg}\,\,{\rm cm}^{-2} 
\,\,{\rm s}^{-1} } \right) \, . 
\label{zetaxnumber} 
\ee 
For the given background X-ray flux value, the X-ray ionization rate
is larger than the cosmic ray ionization rate by a factor of $\sim4-8$
(see also Lorenzani \& Palla 2001).  Keep in mind, however, that the
cosmic ray ionization rate varies with galactocentric radius and with
column density of the region (van der Tak \& van Dishoeck 2000).
Likewise, the X-ray fluxes have a broad distribution of values (as
shown in Figures \ref{fig:fluxdist} -- \ref{fig:flux_compbig}), so
that the X-ray ionization rate can be much larger, or smaller, than
the benchmark value given in equation (\ref{zetaxnumber}).  The
relative importance of cosmic ray and X-rays for ionization is thus
expected to vary significantly from cluster to cluster. 

The cluster is essentially optically thin to X-rays, so that young
stellar objects within the cluster receive the unattenuated flux.  
We can estimate the optical depth as follows: The cross section of
equation (\ref{csection}) can be integrated over the energy range of
the observational surveys (the energy range used to construct the flux
distributions) to find an average value $\langle \sigma \rangle$
$\approx$ $\sigma_0/10 \approx 2 \times 10^{-23}$ cm$^2$. The number
density $n$ of a cluster with $N$ = 100 -- 1000 and radius $R_c\sim1$
pc is about $n$ = 3000 cm$^{-3}$. The mean free path of an X-ray is
thus $\ell=1/n\langle\sigma\rangle\approx5{\rm pc}>R_c$.

Another way to gauge the importance of the X-ray radiation fields is
to compare the flux levels to those of the background galaxy and the
background universe.  Here we are interested in X-rays in the range
0.2 -- 15 keV, corresponding to the range in the observational surveys
used to construct the X-ray luminosity vs mass relations used in this
paper.  The total flux in soft X-rays (in the range 0.28 -- 1 keV) is
estimated to lie in the range $f_X \approx 5 - 10 \times 10^{-8}$ erg
cm$^{-2}$ s$^{-1}$ (Fried et al. 1980), where we have assumed an
isotropic radiation field to convert (energy-integrated) specific
intensity to flux. This value has considerable uncertainty. In
addition, comparable fluxes are present in harder X-rays, e.g., the
range 1 -- 15 keV that provides the other half of the range under
consideration. These harder X-rays are thought to come from
extragalactic sources (and are observed to be nearly isotropic),
whereas the softer X-rays have a significant Galactic contribution
(e.g., Boldt 1987). In any case, the total X-ray background flux is
about $f_X \approx 1-2 \times 10^{-7}$ erg cm$^{-2}$ s$^{-1}$. Given
the distributions of cluster properties considered here, and the
distributions of fluxes for a given type of cluster, the X-ray flux
provided by clusters $F_{XC}$ is larger than the background $f_X$ by
factors in the range 10 -- 1000. 

\section{Conclusion} 

This paper constructs the expected X-ray radiation fields provided by
young embedded clusters. Specifically, we calculate the distributions
of X-ray luminosity for clusters of a given membership size $N$ and
the corresponding distributions of X-ray flux. We also determine the
distributions of X-ray flux for two cluster ensembles (for the cluster
distribution found in the solar neighborhood and for an extended
distribution with maximum membership size $N_{max}=10^5$).  The flux
distributions depend on the density profile of the stellar objects
(the individual X-ray sources) within the cluster and we consider
models that span the expected range of possibilities. The main results
can be summarized as follows:

The expected flux levels for X-rays are modest, with a characteristic
value of order $F_X \sim 1-6 \times 10^{-5}$ erg cm$^{-2}$ s$^{-1}$;
the range corresponds to the types of clusters under consideration,
where the X-ray flux increases with cluster membership size $N$ and
with the central concentration of the density profile (Figures
\ref{fig:fluxdist} -- \ref{fig:flux_compbig}).  The distributions of
X-ray flux are relatively wide, however, with the full width at
half-maximum for $\log_{10} F_X$ corresponding to a factor of
$\sim3-4$ for $\rho\propto 1/r$ density profiles and a factor of
$\sim5-10$ for $\rho\propto 1/r^2$ density profiles (see Figures
\ref{fig:fluxdist} and \ref{fig:fluxdistalt}). The cluster to cluster
variation in X-ray flux will thus be significant.  For completeness,
we also note that X-ray emission is highly variable with time, and
that the scatter in X-ray luminosity generally decreases with cluster
age (Alexander \& Preibisch 2012).

Given the expected X-ray flux distributions for embedded clusters, as
calculated herein, the outer regions of circumstellar disks (where
planets form) receive comparable amounts of flux from their central
stars and from the background. The central star dominates for small
clusters if one ignores geometric effects (equation [\ref{fxest}]),
whereas the background X-ray flux from the cluster dominates for
geometrically thin disks (equation [\ref{xraystar}]) and also for
larger clusters (Figure \ref{fig:flux_compbig}).  Circumstellar disks
are expected to be relatively thin and the flux impinging on them from
the central star is significantly reduced by geometric effects. As a
result, the background X-ray flux is expected to exceed that of the
star for disk radii beyond $r\approx 9-14$ AU. We stress that
significant variability in these values are expected given both the
large variability in the X-ray luminosities of individual stars and
the broad X-ray flux distributions within the cluster environment.

Whereas disks receive comparable X-ray fluxes from their central stars
and the background cluster, the situation is markedly different for
the case of both FUV and EUV radiation. At UV wavelengths, the
background radiation fields of the cluster overwhelm those of the
central stars for clusters with $N \ge 100$ (Armitage 2000; Adams \&
Myers 2001; Fatuzzo \& Adams 2008). As a result, photoevaporation of
circumstellar disks due to UV radiation is often dominated by the
cluster background (not the central star), whereas mass loss due to
X-rays can be dominated by the central star (and not the cluster).

The ionization rate provided by X-rays in the cluster environment has
a characteristic value $\zeta_X \sim 8 \times 10^{-17}$ s$^{-1}$ (see
equation [\ref{zetaxnumber}]). This ionization rate is about 4 -- 8
times the fiducial ionization rate $\zeta_{CR}$ from cosmic rays,
which are often considered to be the dominant source of ionization for
molecular clouds and star formation (e.g., Shu 1992). The ionization
rates for both X-rays $\zeta_X$ and cosmic rays $\zeta_{CR}$ have wide
distributions, so that the relative importance of the two sources will
vary significantly from cluster to cluster (and can vary with time).
In a similar vein, the background of X-ray radiation produced by the
cluster is larger than the background radiation provided by the galaxy
or the universe. Over the energy range under consideration, the
cluster contribution dominates by a factor in the range 10 -- 1000,
with a typical value of $\sim100$.

The X-ray background radiation in young embedded clusters has a number
of potential effects on star and planet formation (Glassgold et al.
2000; Feigelson et al.  2007).  The elevated levels of ionization lead
to greater coupling between the cluster gas and magnetic fields (e.g.,
Shu 1992); this increased coupling, in turn, leads to longer time
scales for ambipolar diffusion and acts to slow down additional star
formation.  Enhanced ionization also affects MHD-driven turbulence,
both in the cluster gas and in circumstellar disks via MRI (Balbus \&
Hawley 1991).  The background X-ray flux will produce line diagnostics
in circumstellar disks (e.g., Tsujimoto et al. 2005; Hollenbach \&
Gorti 2009) and will affect chemical reactions in the gas (Aikawa \&
Herbst 1999). These processes, and many others, should be studied in
the future to provide a more complete picture of how cluster
environments shape the solar systems forming within them.

\vskip 0.35truein 
\centerline{\bf Acknowledgment} 
We thank the referee for their careful review of the manuscript, and 
for their useful comments that improved the quality of the writing.  
FCA is supported at UM by NASA grant NNX11AK87G from the Origins of
Solar Systems Program, and by NSF grant DMS-0806756 from the Division
of Applied Mathematics. MF is supported at XU through the Hauck
Foundation. LH is supported at NKU through a CINSAM Research Grant.

%\newpage 


\bigskip 
\begin{thebibliography} 
\medskip 

\bibitem[Adams(2010)]{adamssun} 
Adams, F. C. 2010, ARA\&A, 48, 47 

\bibitem[ahlg]{ahlg} 
Adams, F. C., Hollenbach, D., Laughlin, G., \& Gorti, U. 
2004, ApJ, 611, 360 

\bibitem[am]{am01}
Adams, F. C., \& Myers, P. C. 2001, ApJ, 553, 744

\bibitem[apfm]{apfm} 
Adams, F. C., Proszkow, E. M., Fatuzzo, M., \& Myers, P. C. 
2006, ApJ, 641, 504 

\bibitem[as]{as86}
Adams, F. C., \& Shu, F. H. 1986, ApJ, 308, 836

\bibitem[aikawa]{aikawa} 
Aikawa, Y., \& Herbst, E. 1999, A\&A, 351, 233 

\bibitem[Alexander \& Preibisch(2012)]{alexpre} 
Alexander, F., \& Preibisch, T. 2012, A\&A, 539, A64

\bibitem[alex04]{alex04}
Alexander, R. D., Clarke, C. J., \& Pringle, J. E. 2004, MNRAS, 354, 71 

\bibitem[alex05]{alex05}
Alexander, R. D., Clarke, C. J., \& Pringle, J. E. 2005, MNRAS, 358, 283 

\bibitem[alex06]{alex06}
Alexander, R. D., Clarke, C. J., \& Pringle, J. E. 2006, MNRAS, 369, 229 

\bibitem[allen]{allen} 
Allen, L. E., Megeath, S. T., Gutermuth, R., Myers, P. C., Adams,
F. C., Muzzerolle, J., Young, E., \& Pipher, J. L. 2007, Protostars 
and Planets V, ed. B. Reipurth (Tucson: Univ. Ariz. Press), p. 361

\bibitem[arm]{arm} 
Armitage, P. J. 2000, A\&A, 362, 968 

\bibitem[bh]{bh}
Balbus, S., \& Hawley, J. 1991, ApJ, 376, 214

%\bibitem[balog]{balog}
%Balog, Z, Muzerolle, J., Rieke, G. H., Su, K. Y. L, \& Young, E. T.
%2007, ApJ, 660,  1532

%\bibitem[beth]{beth} 
%Bethell, T. J., Zweibel, E. G., \& Li, P. S. 2007, ApJ, 667, 275 

\bibitem[Boldt(1987)]{boldt} 
Boldt, E. 1987, Phys. Reports, 146, 215 

\bibitem[carp]{carp00}
Carpenter, J. M. 2000, AJ, 120, 3139

\bibitem[chandar]{chandar}
Chandar, R., Bianchi, L., \& Ford, H. C. 1999, ApJS, 122, 431

%\bibitem[chab]{chab}
%Chabrier, G. 2003, PASP, 115, 763 

%\bibitem[clark]{clark}
%Clarke, C. J., Gendrin, A., Sotomayor, M. 2001, MNRAS, 328, 485 

%\bibitem[ee]{e2}
%Elmegreen, B. G., Efremov, Y., Pudritz, R. E., \& Zinnecker, H. 2000,
%Protostars and Planets IV, eds. V. Mannings, A. Boss, \& S. Russell  
%(Tucson: Univ. Ariz.  Press), p. 179

\bibitem[erc(2008)]{erc08} 
Ercolano, B., Drake, J. J., Raymond, J. C., \& Clarke, C. J. 2008, ApJ, 688, 398 

\bibitem[erc(2009)]{erc09} 
Ercolano, B., Clarke, C. J., \& Drake, J. J. 2009, ApJ, 699, 1639 

\bibitem[fa(3008)]{fa} 
Fatuzzo,  M., \& Adams, F. C. 2008, ApJ, 675, 1361

\bibitem[fe(93)]{fe93} 
Feigelson, E. D., Casanova, S., Montmerle, T., \& Guibert, J. 1993, ApJ, 416, 623 

\bibitem[fe(02)]{fe02} 
Feigelson, E. D., Broos, P., Gaffney, J. A. III, Garmire, G.,
Hillenbrand, L. A., Pravdo, S. H., Townsley, L., \& Tsuboi, Y. 2002,
ApJ, 574, 258

\bibitem[fe(05)]{fe05} 
Feigelson, E. D., Getman, K.,  Townsley, L., Garmire, G., Preibisch, T.,
Grosso, N., Montmerle, T., Muench, A., \& McCaughrean, M.  2005,
ApJS, 160, 379

\bibitem[fe(07)]{fe07} 
Feigelson, E. D., Townsley, L., G{\"u}del, M., \& Stassun, K.  2007,
Protostars and Planets V, ed. B. Reipurth (Tucson: Univ. Ariz. Press),
p. 313

\bibitem[Fried et al.(1980)]{fried} 
Fried, P. M., Nousek, J. A., Sanders, W. T., \& Kraushaar, W. L. 
1980, ApJ, 242, 987 

%\bibitem[gam]{gam}
%Gammie, C. F. 1996, ApJ, 457, 355 

\bibitem[Glassgold et al.(1997)]{glass} 
Glassgold, A. E., Najita, N., \& Igea, J. 1997, ApJ, 480, 344 

\bibitem[Glassgold et al.(2000)]{glassppiv} 
Glassgold, A. E., Feigelson, E. D., \& Montmerle, T. 2000, 
Protostars and Planets IV, eds. V. Mannings, A. Boss, \& S. Russell  
(Tucson: Univ. Ariz.  Press), p. 429

%\bibitem[gpm]{gpm} 
%Guarcello, M. G., Prisinzano, L., Micela, G., Damiani, F., Peres, G.,
%& Sciortino, S. 2007, A\&A, 462, 245

\bibitem[gh]{gh} 
Gorti, U., \& Hollenbach, D. 2002, ApJ, 573, 215 

\bibitem[gh2]{gh2} 
Gorti, U., \& Hollenbach, D. 2009, ApJ, 690, 1539 

%\bibitem[gou]{gou} 
%Gounelle, M., \& Meibom, A. 2007, submitted to ApJ 

%\bibitem[haisch]{haisch}
%Haisch, K. E., Lada, E. A., \& Lada, C. J. 2001, ApJ, 553, L153

%\bibitem[hq]{hq} 
%Hernquist, L. 1990, ApJ, 356, 359 

\bibitem[holden]{holden}
Holden, L., Landis, E., Spitzig, J., \& Adams, F. C. 2011,  PASP, 123, 14

\bibitem[hg]{hg} 
Hollenbach, D., \& Gorti, U. 2009, ApJ, 703, 1203 

\bibitem[h94]{h94}
Hollenbach, D., Johnstone, D., Lizano, S., \& Shu, F. 1994, ApJ, 654, 669 

\bibitem[hyj]{hyj} 
Hollenbach, D. J., Yorke, H. W., \& Johnstone, D. 2000, Protostars and
Planets IV, eds. V. Mannings, A. Boss, \& S, Russell (Tucson: Univ. 
Ariz.  Press), p. 401

%\bibitem[ki]{ki} 
%Kobayashi, H., \& Ida, S. 2001, Icarus, 153, 416 

\bibitem[Krolik \& Kallman(1983)]{krolik} 
Krolik, J. H., \& Kallman, T. R. 1983, ApJ, 267, 610 

\bibitem[kr]{kr}
Kroupa, P. 2001, MNRAS, 322, 231

%\bibitem[kr]{kr}
%Kroupa, P. 2002, Science, 295, 82 

\bibitem[ll]{ll}
Lada, C. J., \& Lada, E. A. 2003, ARA\&A, 41, 57

\bibitem[lorenpalla]{lorenpalla} 
Lorenzani, A., \& Palla, F. 2001, in ASP Conf. Ser. 243, Darkness to
Light, ed. T. Montmerle \& Ph. Andr{\'e} (San Francisco: ASP), 745

%\bibitem[laugh]{laugh}
%Laughlin, G., Bodenheimer, P., \& Adams, F. C. 2004, ApJ, 612, 73 

%\bibitem[]{mm87}
%Maeder, A., \& Meynet, G. 1987, A\&A, 182, 243 

%\bibitem[malm]{malm} 
%Malmberg, D., de Angeli, F., Davies, M. B., Church, R. P., Mackey, D.,
%\& Wilkinson, M. I.  2007, MNRAS, 378, 1207

%\bibitem[mass]{mass}
%Massey, P. 2003, ARA\&A, 41, 15 

%\bibitem[ops]{ops} 
%Olczak, C., Pfalzner, S., \& Spurzem, R. 2006, ApJ, 642, 1140

%\bibitem[adh]{adh} 
%Ouellette, N., Desch, S. J., Hester, J. J. 2007, ApJ, 662, 1268O

%\bibitem[phm]{phm} 
%Parravano, A., Hollenbach, D. J., \& McKee, C. F. 2003, ApJ, 584, 797 

%\bibitem[poe]{poe} 
%Pfalzner, S., Olczak, C., \& Eckart, A. 2006, A\&A, 454, 811

\bibitem[pfalzner]{pfalzner} 
Pfalzner, S. 2009, A\&A, 498, 37

\bibitem[porras]{porras}
Porras, A., et al. 2003, AJ, 126, 1916

\bibitem[preb]{preibisch}
Preibisch, T., Kim, Y.-C., Favata, F., Feigelson, E. D., Flaccomio,
E., Getman, K., Micela, G., Sciortino, S., Stassun, K., Stelzer, B.,
\& Zinnecker, H. 2005, ApJS, 160, 401

\bibitem[proszkow]{proszkow}
Proszkow, E.-M., \& Adams, F. C. 2009, ApJS, 185, 486 

\bibitem[rich]{rich}
Richtmyer, R. D. 1978, Principles of Advanced Mathematical Physics 
(New York: Springer) 

%\bibitem[]{salp55} 
%Salpeter, E. E. 1955, ApJ, 121, 161 

%\bibitem[]{sc01} 
%Scally, A., \& Clarke, C. 2001, MNRAS, 325, 449 

%\bibitem[]{ssmm92} 
%Schaller, G., Schaerer, D., Meynet, G., \& Maeder, A. 1992, 
%A\&AS, 96, 269  

\bibitem[shu92]{shu92}
Shu, F. H. 1992, Gas Dynamics, (Mill Valley: Univ. Sci. Books)  

%\bibitem[sal]{sal} 
%Shu, F. H., Adams, F. C., \& Lizano, S. 1987, ARA\&A, 25, 23 

\bibitem[]{sjh}
Shu, F. H., Johnstone, D., \& Hollenbach, D. J. 1993, Icarus, 106, 92 

\bibitem[]{sh99}
St\"orzer, H., \& Hollenbach, D. 1999, ApJ, 515, 688

%\bibitem[]{tpn}
%Testi, L., Palla, F., \& Natta, A. 1998, A\&AS, 133, 81

%\bibitem[tb]{tb}
%Throop, H. B., \& Bally, J. 2005, ApJ, 623, 149 

\bibitem[tsuji]{tsuji} 
Tsujimoto, M., Feigelson, E. D., Grosso, N., Micela, G., Tsuboi, Y.,
Favata, F., Shang, H., \& Kastner, J. H. 2005, ApJS, 160, 503 

\bibitem[vv]{vv} 
van der Tak, F.F.S., \& van Dishoeck, E. F. 2000, A\&A, 358, L79 

%\bibitem[wg]{wg} 
%Williams, J. P., \& Gaidos, E. 2007, ApJ, 663, 33

%\bibitem[Yan et al.(1998)]{yan} 
%Yan, M., Sadeghpour, H. R., \& Dalgarno, A. 1998, ApJ, 496, 1044 

%\bibitem[zman]{zman} 
%Zahnle, K., Arndt, N., Cockell, C., Halliday, A., Nisbet, E., Selsis,
%F., \& Sleep, N. H.  2007, SSRv, 129, 35Z


\end{thebibliography}
\end{document}